\documentclass{article}
\usepackage[T1]{fontenc}
\usepackage[utf8]{inputenc}
\usepackage{ismir} % Remove the "submission" option for camera-ready version
\renewcommand{\permission}{}  % clear footer for arxiv
\usepackage{fancyhdr}
\usepackage{amsmath,cite,url}
\usepackage{titlesec}
\usepackage{amssymb}
\usepackage{graphicx}
\usepackage{color}
\usepackage{booktabs}
\usepackage{siunitx}
\usepackage{etoolbox}
\usepackage{enumitem}
\usepackage{caption}
\usepackage{subcaption}
\usepackage{microtype}           % better kerning & hyphenation for arXiv
\usepackage{tcolorbox}
\tcbuselibrary{skins}
\robustify\bfseries
\robustify\underline

\usepackage{array}
\usepackage[table]{xcolor}

\definecolor{secblue}{HTML}{1F3A5F}      % deep navy for headings & labels
\definecolor{tldrblue}{HTML}{2C5282}     % slightly brighter navy for TL;DR tab
\definecolor{rowtint}{HTML}{F2F4F7}      % very light gray for zebra stripes
\definecolor{melcolor}{HTML}{4472C4}
\definecolor{rhycolor}{HTML}{E07B39}
\definecolor{timcolor}{HTML}{47A25A}

\tcbset{
  tldr/.style={
    enhanced,
    colback=tldrblue!4,
    colframe=tldrblue!85,
    boxrule=0.6pt,
    arc=5pt,
    left=8pt, right=8pt, top=5pt, bottom=5pt,
    title={\bfseries\small In a Nutshell},
    fonttitle=\bfseries,
    coltitle=white,
    attach boxed title to top left={yshift=-2.2mm, xshift=8pt},
    boxed title style={colback=tldrblue, arc=2pt, boxrule=0pt},
    before skip=0pt, after skip=10pt,
    fontupper=\small
  }
}

\titleformat{\section}
  {\normalfont\large\bfseries\color{secblue}}
  {\thesection.}{0.6em}{}
\titleformat{\subsection}
  {\normalfont\normalsize\bfseries\color{secblue}}
  {\thesubsection}{0.6em}{}
\titleformat{\subsubsection}
  {\normalfont\small\bfseries\itshape\color{secblue!85!black}}
  {\thesubsubsection}{0.6em}{}

\title{MERIT: Learning Disentangled Music Representations for Audio Similarity}

\oneauthor
  {Abhinaba Roy, Junyi Liang, and Dorien Herremans}
  {Singapore University of Technology and Design \\
  \texttt{abhinaba\_roy@sutd.edu.sg, liangjunyi010@gmail.com} \\ 
  \texttt{dorien\_herremans@sutd.edu.sg}}
\usepackage[bookmarks=false,
            colorlinks=true,
            linkcolor=blue!60!black,
            citecolor=blue!55!black,
            urlcolor=teal!80!black,
            pdftitle={MERIT: Learning Disentangled Music Representations for Audio Similarity},
            pdfauthor={Abhinaba Roy, Junyi Liang, and Dorien Herremans}]{hyperref}
            
\begin{document}

\maketitle

% ── TL;DR ──────────────────────────────────────────────────────────────────
\begin{tcolorbox}[tldr]
\textbf{MERIT} learns three lightweight projection heads
(melody, rhythm, timbre) on top of a pre-trained \emph{frozen} MERT backbone.
A generative triplet pipeline provides factor-isolated training data with no manual labelling.
Each head achieves $\geq\!99.6\%$ triplet accuracy on held-out factor-controlled tests and
the correct head dominates on all three independent zero-shot audio probes,
confirming that disentanglement transfers to real-world audio.
\end{tcolorbox}

\begin{abstract}
Current music similarity models typically compute a single, monolithic score, entangling distinct musical dimensions like melody, rhythm, and timbre. This limits user control and interpretability, making it impossible to execute nuanced queries. We introduce MERIT, a framework for learning disentangled, factor-specific music representations tailored to these three core dimensions. To overcome the lack of isolated musical variations in real-world audio, we use a novel training strategy that uses conditional audio generation and source-separated stems to strongly encourage single-factor variation in training data. Our evaluations demonstrate strong factor-wise disentanglement. Each head responds strongly to its intended perceptual dimension while remaining near chance on the others, a representational property that holds across both the synthetic training domain and independent real-world audio.
\end{abstract}
%\section{Keywords}  
  %Video Question Answering. Music-Video Causality. Multimodal Reasoning. Automated Dataset Generation.

\section{Introduction}
\label{sec:intro}

Music similarity is inherently multi-dimensional. A solo piano cover of a rock anthem preserves the melody and harmonic identity of the original while replacing every instrument and reshaping the groove. Two recordings by the same artist often share a timbral signature with no constraint at all on melody. Within a dance genre, different tracks might share the same syncopation pattern despite covering different songs and using different instrumentation. Listeners can attend to any one of these dimensions independently, use it as the basis of a similarity judgement, and articulate why two recordings are alike along that axis without committing to a single overall verdict.

Most audio similarity systems collapse this structure into singular similarity number. Learned audio embeddings such as CLAP~\cite{wu2023large} and MuLan~\cite{huang2022mulan}, and self-supervised audio backbones such as MERT~\cite{li2023mert}, expose a single vector per clip; cosine similarity in that vector is an aggregate score whose dependence on melody, rhythm, and timbre is fixed by pre-training and not exposed to the user. Factor-specific specialists exist for cover song detection~\cite{yu2020learning}, melody similarity~\cite{lu2025melodysim}, and instrument classification, caption generation~\cite{chopra2025sonicverse} but each addresses one dimension in isolation and cannot expose the others as comparable scores from the same query.
To the authors' knowledge, no existing framework learns three similarity spaces from a unified pipeline such that each space is shaped by a different perceptual factor and all three can be queried together.

\begin{figure}[h!]
  \centering
  \includegraphics[width=0.9\columnwidth, trim={0 14px 0 10px}, % Left Bottom Right Top
  clip]{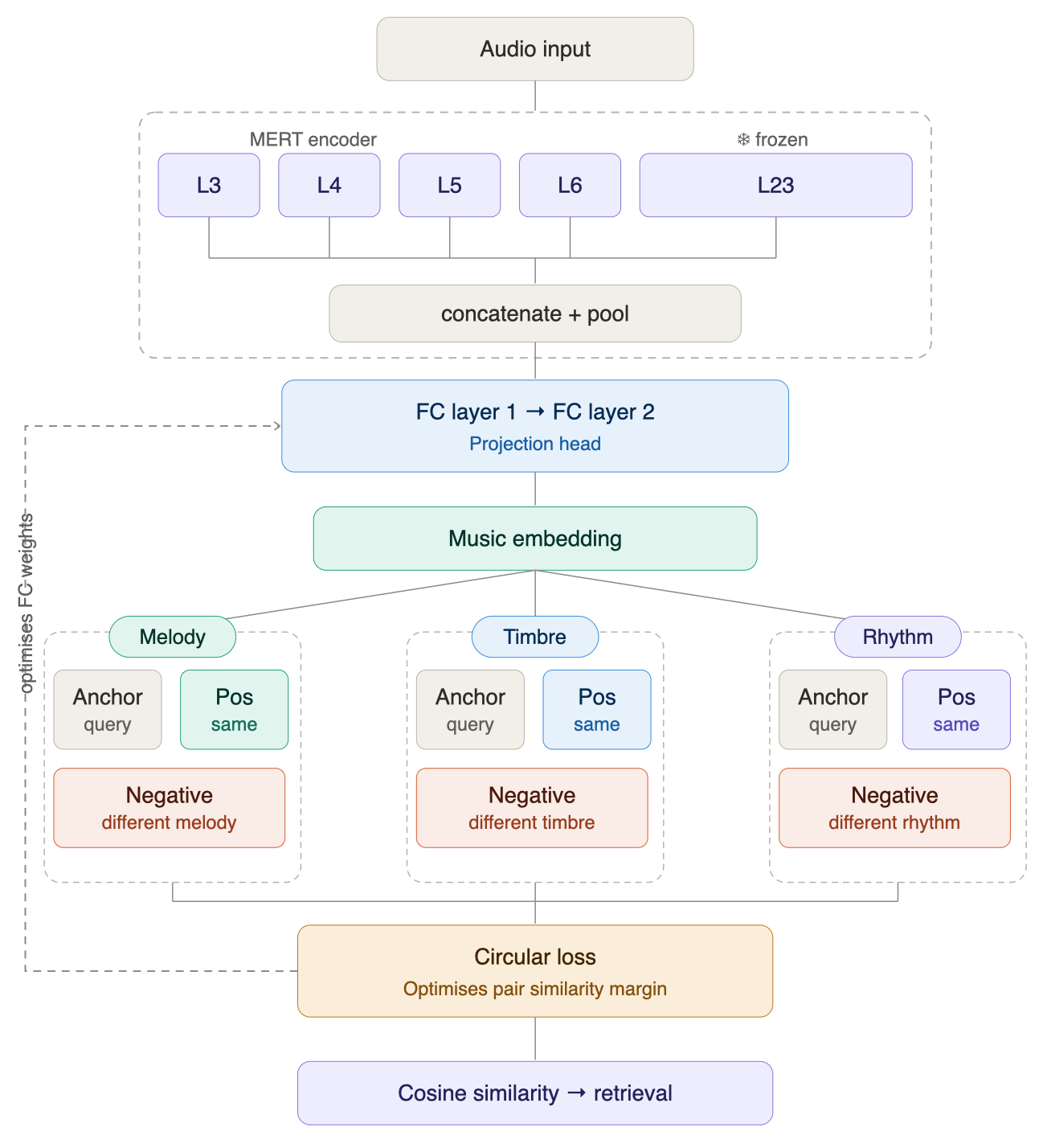}
  \caption{\textbf{MERIT architecture and training pipeline.}
    \emph{Backbone:} Audio input passes through frozen MERT encoder, layers 3-6, and~23 are concatenated and pooled.
    \emph{Projection:} Three \textbf{distinct \& separate} projection head maps the pooled features into Melody, Timbre, and Rhythm. 
    \emph{Optimization \& Inference:} A circular loss maximizes the pair similarity margin across all triplets. During inference, retrieval is performed by cosine similarity.}
  \label{fig:overview}
  % \label{fig:overview}
\end{figure}

We present \textbf{MERIT}, a representational framework that learns three independent projection heads (one each for \emph{melody}, \emph{rhythm}, and \emph{timbre}) from a frozen encoder backbone  (Fig.~\ref{fig:overview}). Rather than optimizing for an aggregated similarity score, our objective is the structural decoupling of these musical dimensions. The primary aim of MERIT is to achieve high functional selectivity, where each projection head responds exclusively to its target factor while remaining invariant to others.

A key challenge in learning such granular representations is the lack of training data where musical factors are strictly isolated. Real-world recordings naturally entangle melody, rhythm, and timbre, making it difficult for a model to learn their independent boundaries. To address this bottleneck, we propose a triplet construction strategy that leverages both conditional audio generation and source separation. This allows us to curate a large-scale training set where only a single factor varies at a time, providing the clear supervision signal necessary for true disentanglement.

\begin{figure}[t]
  \centering
  \includegraphics[width=\linewidth]{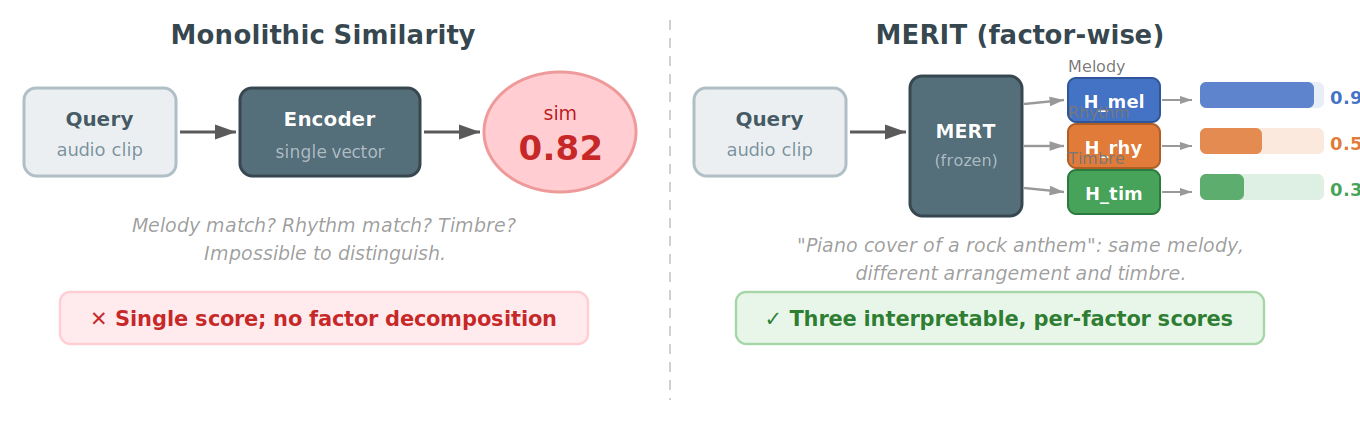}
  \caption{\textbf{Monolithic vs.\ factor-wise music similarity.}
    (\emph{Left})~A conventional encoder collapses melody, rhythm, and timbre into a
    single indistinguishable score.
    (\emph{Right})~MERIT exposes three independent, interpretable scores from the same
    query, enabling targeted retrieval and factor-level explanations.
    Scores are schematic; the scenario illustrates the piano-cover example
    discussed in \S\ref{sec:intro}.}
  \label{fig:concept}
\end{figure}
Each factor head leverages a MERT backbone to capture a multi-scale hierarchy of features. These features are projected through a shallow MLP architecture and trained using Circle Loss to ensure stable and efficient representation learning. Because the backbone is frozen, training is fast while maintaining the robust musical knowledge of the underlying foundation model.

We assess the selectivity of MERIT representations using a disentanglement table: each of our three trained heads is evaluated on all three factor test sets. We also evaluate standard, monolithic audio representations—including CLAP and the raw MERT backbone on these same factor-controlled tests.
Beyond this internal evaluation, we subject the model to zero-shot diagnostic probes using real-world collections with known perceptual structures, such as rhythmic style datasets for groove. The goal is to determine if the cross-factor selectivity pattern observed in training holds true across diverse, non-synthetic audio environments.

In summary, the main contributions of this paper are:
\begin{enumerate}[itemsep=1mm, parsep=0pt]
  \item A scalable data pipeline for constructing factor-controlled music triplets via generative conditioning and source separation, along with our constructed dataset. 
  \item MERIT, a representational architecture that demonstrates high functional selectivity by decoupling entangled musical dimensions into independent, addressable scoring channels.
  \item An evaluation protocol that quantifies factor-wise selectivity, alongside zero-shot probes confirming that this selectivity generalizes to independent, real-world audio collections.
\end{enumerate}
Code and pre-trained models are available at \url{https://github.com/AMAAI-Lab/MERIT}.

\section{Related Work}

\textbf{General audio and music embeddings.}
Large-scale contrastive audio--language pre-training, as in CLAP~\cite{wu2023large} and MuLan~\cite{huang2022mulan}, produces rich audio representations by aligning audio with free-form text descriptions.
Self-supervised music encoders such as MERT~\cite{li2023mert} extend masked language modelling to audio with auxiliary pitch, chroma, and beat objectives, yielding representations that probe well on a wide range of music understanding tasks. Such systems produce a single embedding per clip whose dependence on melody, rhythm, and timbre is entangled and fixed by pre-training.

\textbf{Contrastive and metric learning for audio.}
Contrastive learning with triplet~\cite{hermans2017defense} and Circle Loss~\cite{sun2020circle} objectives has been applied to audio fingerprinting~\cite{chang2021neural}, speaker verification~\cite{zhang2021contrastive}, and music tagging~\cite{spijkervet2021contrastive,meseguer2024experimental}.
Most of these applications define a single similarity space; our work extends the paradigm to three simultaneous factor-specific spaces trained on independent, factor-controlled triplet datasets.
Luo~et~al.~\cite{luo2019learning} learn disentangled timbre and pitch representations for musical instrument sounds using Gaussian mixture VAEs, motivating the general principle that musical factors can be separated in the latent space; MERIT scales this idea to full musical recordings via a discriminative, shared backbone approach.

\textbf{Factor-specific retrieval.}
Cover song detection~\cite{yu2020learning} focuses narrowly on melodic and harmonic identity, using features such as chroma, key, and harmonic pitch-class profiles.
MelodySim~\cite{lu2025melodysim} specifically targets melodic similarity for short audio clips.
Beat and rhythm characterisation and instrument classification address complementary factors in isolation.

% None of these specialist models exposes the other factors as comparable scores, making it impossible to express a multi-factor query or to interpret why a retrieved result was ranked highly. 
MERIT is, to our knowledge, the first framework that learns all three of melody, rhythm, and timbre similarities simultaneously, while leveraging a foundational embedding, and returns them as interpretable scores.

% Somewhat relevant paper of mine \cite{luo2019learning} also accepted in ISMIR -> https://arxiv.org/abs/1906.08152  Learning Disentangled Representations of Timbre and Pitch for Musical Instrument Sounds Using Gaussian Mixture Variational Autoencoders

% MelodySim \cite{lu2025melodysim}

\section{Method}

\subsection{Factor-Specific Triplet Construction}
\label{sec:triplets}

A training triplet for factor $f$ is a tuple $(A,\, P_f,\, N)$ where anchor~$A$ and positive~$P_f$ are similar on factor $f$ and differ in other respects, while negative~$N$ differs from~$A$ on factor $f$.
We construct three separate triplet datasets, one per factor, using different conditioning strategies.

Given $k$ positives per anchor, we expand each folder into $k^2$ valid triplets: the $k$ anchor--positive triplets $(A, P_i, N)$ plus all $k(k-1)$ cross-positive triplets $(P_i, P_j, N)$ with $i \neq j$.
This is valid because all positives in a folder share the exact same factor-defining property, making any positive a legitimate anchor for any other.

\subsubsection{Melody Triplets}
For each anchor~$A$ drawn from a melodic stem in MoisesDB~\cite{pereira2023moisesdb}, we compute a pitch salience map via probabilistic YIN (pYIN) F0 estimation, producing a one-hot pitch matrix at 50~frames/s over 53 MIDI bins (G2--B6).
This map is passed as a melodic conditioning signal to JASCO~\cite{tal2024joint}, a music generation model that supports conditioning on pitch content.
The generated positive $P_\text{mel}$ follows the melodic contour of~$A$ while instrumentation and rhythmic pattern are determined by a text prompt sampled uniformly at random from a bank of 5{,}000 diverse style descriptions spanning genres, instruments, and moods (e.g.\ ``Folk song with accordion and acoustic guitar'', ``Slow blues rock with electric guitar riffs and steady drums'').  

%Randomising the text prompt across this diverse bank ensures that no single instrumentation or genre is consistently paired with any anchor, forcing the head to rely on melodic contour rather than timbral or stylistic shortcuts.
A negative~$N$ is any recording outside the same anchor folder.
We generate five positives per anchor, yielding 5{,}000 folders and \textbf{125{,}000 training triplets} after $k^2$ expansion.

\subsubsection{Rhythm Triplets}
We follow the same pipeline but replace the pitch salience conditioning with the drum stem of~$A$, which MoisesDB provides as a pre-separated stem.
JASCO conditioned on a drum stem preserves the anchor's temporal groove and beat pattern while the pitched melodic content is controlled by a text prompt drawn from the same 5{,}000-prompt bank used for melody triplets, again sampled independently for each generated clip.
The positive $P_\text{rhy}$ therefore shares the anchor's rhythmic structure but may have an entirely different melody, key, and instrumentation.
This dataset also yields 5{,}000 folders and \textbf{125{,}000 training triplets}.

\subsubsection{Timbre Triplets}
Timbral similarity is operationalized as instrument-class identity.
From MoisesDB, we extract source-separated stems annotated with instrument labels.
Anchor~$A$ and positive~$P_\text{tim}$ are stems from two \emph{different} songs assigned the same instrument label (e.g., both are piano stems); the negative~$N$ is a stem with a different instrument label drawn from the \emph{same song} as~$A$.
No generative model is required.
With up to five positives per anchor, this dataset yields 1{,}855 folders and \textbf{46{,}241 training triplets} after $k^2$ expansion.
All datasets are split at the folder level (90\%~train / 10\%~test) with a fixed random seed.
The full dataset is available at \url{https://huggingface.co/datasets/amaai-lab/merit} under CC BY-NC-SA~4.0.

\subsection{Frozen MERT Multi-Layer Backbone}

All three factor heads start with a frozen MERT-v1-330M~\cite{li2023mert} encoder. This is a 330-million-parameter masked audio language model pre-trained on approximately 160k hours of music with auxiliary pitch, chroma, and beat prediction objectives.
Rather than relying on the final hidden state alone, we extract activations from five transformer layers: layers 3, 4, 5, 6, and 23 ~\cite{liu2024leveraging}.
Each layer yields a $T \times 1024$ hidden state sequence, which we mean-pool over the time axis; the five resulting 1024-dimensional vectors are concatenated into a single \textbf{5120-dimensional} representation per clip.

This multi-scale representation captures complementary levels of musical abstraction: \textbf{Layers 3--6} encode low- and mid-level acoustic structure. \textbf{Layer 23} (penultimate) encodes high-level semantic content.

Using a common backbone for all three factors eliminates the encoder as a confounding variable. Differences in disentanglement quality are attributable solely to the training data and projection head, not to architectural differences.
%All MERT weights are frozen throughout; the three projection heads are the \emph{only} trainable parameters in MERIT.

\subsection{Trainable Projection Heads}

Each factor head $h_f$ is a shallow two-layer MLP followed by $\ell_2$-normalization:
\begin{equation}
  h_f(\mathbf{z}) = \ell_2\!\left(\mathbf{W}_2^{f}\,\sigma\!\left(\mathbf{W}_1^{f}\,\mathbf{z}\right)\right),
\end{equation}
where $\mathbf{z} \in \mathbb{R}^{5120}$ is the MERT multi-layer embedding,
$\mathbf{W}_1^{f} \in \mathbb{R}^{512 \times 5120}$,
$\mathbf{W}_2^{f} \in \mathbb{R}^{128 \times 512}$ (no output bias), and $\sigma$ denotes ReLU.
The output is a 128-dimensional unit vector; factor similarity between two clips is their cosine similarity in this projected space.

The three heads $H_\text{mel}$, $H_\text{rhy}$, and $H_\text{tim}$ are trained independently using Circle Loss~\cite{sun2020circle} with $\gamma=10$ and margin $m=0.2$:
\begin{equation}
  \mathcal{L} = \operatorname{softplus}\!\bigl(\gamma\,[\alpha_n(S_n - m) - \alpha_p(O_p - S_p)]\bigr),
\end{equation}
where $S_p = \cos(A, P_f)$, $S_n = \cos(A, N)$, $O_p = 1-m$, $\alpha_p = \max(0, O_p - S_p)$, and $\alpha_n = \max(0, S_n - m)$.
Circle Loss re-weights each pair by its current similarity, maintaining gradients on hard pairs that standard triplet loss tends to saturate.
Because the MERT encoder is frozen, embeddings are pre-computed once and stored; head training operates solely on these cached vectors.

\subsection{Retrieval and Score Combination}

At inference time, a query clip is encoded once with the shared frozen MERT backbone to obtain $\mathbf{z}_q \in \mathbb{R}^{5120}$.
Each factor head projects this vector to the 128-dimensional unit sphere:
$\mathbf{q}_f = h_f(\mathbf{z}_q)$.

A reference library is indexed offline using FAISS~\cite{douze2025faiss}: three approximate nearest-neighbor indexes, one per factor, each storing reference clips projected through the corresponding head.
At query time, we score each candidate with all three heads and retrieve the top-10 candidates per factor index.

%, and rank by a weighted combination:
% \begin{equation}
%   S_\text{total}(x) = w_\text{mel}\,S_\text{mel}(x) + w_\text{rhy}\,S_\text{rhy}(x) + w_\text{tim}\,S_\text{tim}(x),
% \end{equation}
% where the default is equal weights ($w_f = \tfrac{1}{3}$) but each weight can be adjusted to reflect the user's intent.
Every returned result exposes its individual $S_\text{mel}$, $S_\text{rhy}$, and $S_\text{tim}$ scores, providing an explicit per-factor explanation for its retrieval.

\section{Experiments and Results}

\subsection{Datasets}
\label{sec:datasets}

All training triplets are derived from MoisesDB~\cite{pereira2023moisesdb}, a multitrack source-separation corpus that provides per-song stems with instrument labels.
Melody and rhythm anchors are MoisesDB stems whose pitch saliency or drum content seeds JASCO~\cite{tal2024joint} to synthesize factor-controlled positives, and timbre triplets are sampled directly from MoisesDB's labelled stems without any generative step (Section~\ref{sec:triplets}).
For zero-shot evaluation we use three external corpora that contain neither JASCO outputs nor MoisesDB material: MUSDB18-HQ~\cite{rafii2019musdb18} provides four pure stem classes (vocals, drums, bass, other) for instrument-class selectivity; we exclude the mixture track because it aggregates several instruments and lacks a single timbral identity.
The Ballroom Dataset~\cite{krebs2013rhythmic} provides eight dance classes defined by rhythmic signature (meter, tempo, syncopation) under near-constant big-band instrumentation.
The Covers80 dataset~\cite{ellis20072007} provides 80 cover song pairs.

\subsection{Training Setup}
Each projection head is trained independently using AdamW optimizer ($\text{lr}=10^{-3}$, weight decay $10^{-4}$, batch size~1024) for 200 epochs with a cosine annealing learning-rate schedule (minimum $\text{lr}=10^{-5}$). As the MERT backbone is frozen, all 5120-dim embeddings are pre-extracted and cached; head training is quite fast. 

\subsection{Human Evaluation of Triplet Quality}
\label{sec:human}

To directly verify that the generated training triplets reflect genuine perceptual distinctions, we conducted a listening test in which participants rated the similarity of audio clip pairs on three 0--100 sliders labelled \emph{melody}, \emph{rhythm}, and \emph{timbre}.
We sampled 10 positive pairs from each of the three training datasets (melody positive pairs, rhythm positive pairs, and timbre positive pairs).
Each pair was rated by 15 music experts.

\begin{table}[t]
\centering
\small
\rowcolors{2}{rowtint}{white}
\begin{tabular}{lccc}
\toprule
\rowcolor{white}
\textbf{Pair type} & \textbf{Melody} & \textbf{Rhythm} & \textbf{Timbre} \\
\midrule
Melody & \textbf{60.0} $\pm$30.3 &  53.4 $\pm$28.8 & 26.3   $\pm$27.5 \\
Rhythm  & 34.0$\pm$27.3 & \textbf{65.8}$\pm$25.6 & 37.5$\pm$26.3 \\
Timbre & 34.2$\pm$27.7 & 37.4$\pm$29.8 & \textbf{57.3} $\pm$31.7\\
\bottomrule
\end{tabular}
\caption{Mean human perceptual similarity ratings (0--100) for each triplet-pair category.
  Standard deviations are given in $\pm$. \textbf{Bold} marks the dominant (intended) factor per row.
  Each cell averages 15 participants $\times$ 30 pairs.}
\label{tab:human}
\end{table}
% Table~\ref{tab:human} shows the ratings.
% Across all three positive-pair types, raters assign the highest score to the dimension that defines the pair---melody positives receive the highest melody rating, rhythm positives the highest rhythm rating, and timbre positives the highest timbre rating---while the off-factor scores remain substantially lower.
% Cronbach's $\alpha$ across raters was TODO for melody, TODO for rhythm, and TODO for timbre, indicating TODO inter-rater agreement.

Table~\ref{tab:human} shows the resulting ratings and confirms that  across all three positive-pair types, raters assign the highest similarity score to the dimension that defines the pair. 
Melody positives receive their highest rating in melody (60.0), rhythm positives in rhythm (65.8), and timbre positives in timbre (57.3). 
While we observe a degree of perceptual coupling; most notably in melody-positive pairs, where rhythmic similarity (53.4) is rated almost as high as melodic similarity (60.0), the intended factor remains the dominant source of perceived similarity in every case. This effect is to be expected as melodies also have a rhythmic aspect to them (e.g.\ note durations).

The internal consistency of these ratings was assessed using Cronbach's $\alpha$. 
We found $\alpha = 0.615$ for melody, $\alpha = 0.493$ for rhythm, and $\alpha = 0.757$ for timbre. 
These values indicate moderately high inter-rater agreement for timbre, acceptable agreement for melody, and moderate agreement for rhythm, suggesting that the perceptual distinctions targeted by our synthetic data are consistently recognizable to human listeners.

\subsection{Internal Disentanglement}
We measure \emph{triplet accuracy} (TA), on held-out test triplets, based on  which of the probe inputs does the model rank closer to the anchor (should be the positive rather then the negative input).
The test set contains 12{,}500 melody triplets, 12{,}500 rhythm triplets, and approximately 4{,}600 timbre triplets, yielding Wald 95\% confidence intervals of about $\pm 0.9$ percentage points on the melody and rhythm cells and $\pm 1.4$ percentage points on the timbre cells.
Table~\ref{tab:disentanglement} reports TA for every combination of model.

\begin{table}[h!]
\centering
\small
\rowcolors{2}{rowtint}{white}
\begin{tabular}{lccc}
\toprule
\rowcolor{white}
\textbf{Model} & \textbf{Melody} & \textbf{Rhythm} & \textbf{Timbre} \\
\midrule
MERT (no head)  & 79.2 & 83.4 & 87.4 \\
CLAP (no head)  & 78.5 & 87.5 & 94.5 \\
\midrule
$H_\text{mel}$ (ours)  & \textbf{99.9} & 58.4 & 60.4 \\
$H_\text{rhy}$ (ours)  & 47.7$^\dagger$ & \textbf{100.0} & 71.6 \\
$H_\text{tim}$ (ours)  & 55.3 & 69.5 & \textbf{99.6} \\
\bottomrule
\end{tabular}
\caption{Triplet accuracy (\%) on held-out factor-controlled test sets.
Rows are models; columns are factor-specific test sets.
Chance = 50\%. \textbf{Bold} entries on the diagonal mark each head's intended factor; $^\dagger$ = below chance.
$N=12{,}500$ per cell on melody/rhythm columns,
$N\!\approx\!4.6\text{k}$ on the timbre column.}
\label{tab:disentanglement}
\end{table}
The three diagonal cells confirm that supervised heads recover their target factor near-perfectly: $H_\text{mel}=99.9\%$ on melody, $H_\text{rhy}=100.0\%$ on rhythm, and $H_\text{tim}=99.6\%$ on timbre, ahead of the strongest unsupervised baseline (CLAP cosine) by 21.4, 12.5, and 5.1 percentage points respectively.
Off-diagonal cells, where each head is asked about a factor it was not trained on, sit far below the diagonal in every case: $H_\text{mel}$ collapses from 99.9\% on melody to 58.4\% on rhythm and 60.4\% on timbre, $H_\text{rhy}$ from 100.0\% on rhythm to 47.7\% on melody and 71.6\% on timbre, and $H_\text{tim}$ from 99.6\% on timbre to 55.3\% on melody and 69.5\% on rhythm.
Baseline encoders show no such pattern: MERT and CLAP cosine remain similarly above chance across all three columns (79.2--87.4\% for MERT, 78.5--94.5\% for CLAP), consistent with a largely holistic representation that is indifferently sensitive to every factor.

The diagonal cells correspond to large positive cosine-distance margins ($d_{AN}-d_{AP}=+0.788$ for $H_\text{mel}$ and $+0.862$ for $H_\text{rhy}$, an order of magnitude above raw MERT's $+0.034$ and $+0.048$ on the same test sets), so the near-saturation reflects a genuinely large effect rather than near-misses being counted as wins.
The most informative entry is the off-diagonal $H_\text{rhy}$ on the melody test set: at 47.7\%, it falls \emph{below} chance, with $d_{AP}=0.667 > d_{AN}=0.628$ and a negative margin of $-0.038$.
% The most informative entry is the off-diagonal $H_\text{rhy}$ on the melody test set: at 47.7\%, it falls \emph{below} chance.
% The mean inter-anchor cosine distances on this cell are $d_{AP}=0.667$ and $d_{AN}=0.628$, giving a negative margin of $-0.038$.
The rhythm head has not merely failed to learn melody; it has learned a geometry in which clips that share a melodic contour sit slightly farther apart than clips that do not, because shared melody in the training distribution co-occurs with mismatched rhythm.
This anti-correlation is strong evidence of disentanglement: the head has actively suppressed a competing factor rather than ignoring it.
% \paragraph{Selectivity gap.}
% We summarise each row of Table~\ref{tab:disentanglement} by a single scalar, the \emph{selectivity gap}
% \[
% \Delta_f = \text{TA}_{f,f} - \tfrac{1}{2}\sum_{g\neq f}\text{TA}_{f,g},
% \]
% the difference between a head's diagonal accuracy and its mean off-diagonal accuracy.
% $\Delta_f \!\approx\! 0$ characterises a holistic representation that is similarly informative on every factor (the behaviour of a single-vector pre-trained encoder); $\Delta_f$ approaching the diagonal-minus-chance ceiling characterises near-ideal disentanglement.
% For MERIT we obtain $\Delta_\text{mel}=+42.3$~pp, $\Delta_\text{rhy}=+40.4$~pp, and $\Delta_\text{tim}=+37.2$~pp, three values in a tight 5-point band that together establish a uniform, factor-agnostic disentanglement profile.
% The corresponding gaps for the unsupervised baselines lie within a few points of zero, since their per-row variance across columns is small.

\subsection{Zero-Shot External Probes}
\label{sec:probes}

To test whether cross-factor selectivity generalises beyond the synthetic training domain, we apply each head, without fine-tuning, to three real-world audio collections that were never seen during training and whose label are independently annotated.
The diagnostic property we look for is which head ranks highest under each label structure, not absolute accuracy on any single benchmark.
Table~\ref{tab:probes} consolidates per-head accuracies; the paragraphs below interpret the cross-factor selectivity pattern.

\begin{table}[t]
\centering
\small
\rowcolors{2}{rowtint}{white}
\begin{tabular}{@{} l c c c c @{}}
\toprule
\rowcolor{white}
\textbf{Probe} & \textbf{MERT} & $H_{\text{mel}}$ & $H_{\text{rhy}}$ & $H_{\text{tim}}$ \\
\midrule
MUSDB18-HQ (timbre)\textsuperscript{*} & \underline{79.8} & 65.4 & 63.1 & \textbf{78.9} \\
Ballroom (rhythm)        & 78.0 & 55.2 & \underline{\textbf{88.0}} & 67.2 \\
Covers80 (covers)        & 66.1 & 63.4 & \underline{\textbf{69.9}} & 61.3 \\
\bottomrule
\end{tabular}
% \vspace{4pt} % Adds a tiny gap between the bottom rule and the footnote
\parbox{\columnwidth}{\scriptsize \textsuperscript{*}Four pure stem classes (vocals, drums, bass, other); the mixture class is excluded.}
\caption{Probing results across different musical facets. \textbf{Bold} marks the best per row among the disentangled heads ($H$); underlined marks the best overall (including the MERT baseline).}
% \vspace{-20 pt}
\label{tab:probes}
\end{table}

% \begin{table}[t]
% \centering
% \caption{Zero-shot triplet accuracy (\%) on three independent real-world audio collections.
% \textbf{Bold} = intended head for each probe; underlined = best individual head per probe.
% The lower block reports the strongest of four score-combination strategies (mean, weighted mean, $\ell_2$-normalised concatenation, element-wise product) per probe.
% $N$ is the number of evaluation triplets per probe.}
% \label{tab:probes}
% \begin{tabular}{lccccc}
% \hline
% \textbf{Probe} & $N$ & MERT raw & $H_\text{mel}$ & $H_\text{rhy}$ & $H_\text{tim}$ \\
% \hline
% MUSDB18-HQ (timbre)$^*$ & 20{,}000 & 79.8 & 65.4 & 63.1 & \underline{\textbf{78.9}} \\
% Ballroom (rhythm)       & 17{,}450 & 78.0 & 55.2 & \underline{\textbf{88.0}} & 67.2 \\
% Covers80 (covers)       & 8{,}100  & 66.1 & 63.4 & \underline{69.9} & 61.3 \\
% \hline
% \multicolumn{6}{l}{\emph{Best score-combination strategy per probe}} \\
% MUSDB18-HQ$^*$         & 20{,}000 & \multicolumn{4}{c}{combined\_max = 79.5; concat = 79.2} \\
% Ballroom               & 17{,}450 & \multicolumn{4}{c}{concat = 80.3; weighted-mean = 79.9} \\
% Covers80               & 8{,}100  & \multicolumn{4}{c}{concat = 69.5; weighted-mean = 69.0} \\
% \hline
% \end{tabular}
% \\[2pt]
% \footnotesize $^*$Four pure stem classes (vocals, drums, bass, other); the ``mixture'' class is excluded because mixture audio is not a single timbral identity.
% \end{table}

\textbf{Probe A. Instrument-class selectivity (MUSDB18-HQ).}
Triplets are constructed from professionally separated multi-instrument stems restricted to the four pure stem classes (vocals, drums, bass, other), with the ``mixture'' track excluded because mixture audio aggregates several instruments and lacks a single timbral identity.
Anchor and positive share an instrument class (e.g., both are drum stems from different songs) and the negative is a stem from a different class drawn from the same anchor song.
$H_\text{tim}$ is the strongest factor head at 78.9\%, while $H_\text{rhy}$ is the weakest at 63.1\%, a 15.8 percentage-point gap on a benchmark whose ground truth is purely timbral.
The raw multi-layer MERT representation scores 79.8\% on this probe, comparable to $H_\text{tim}$, which is unsurprising: instrument identity is a salient axis of MERT's pre-training and survives mean-pooling without supervision.
The novelty here is not that $H_\text{tim}$ exceeds raw MERT but that the other two heads, trained on the \emph{same} backbone, have actively shed instrument-class information relative to it.

\textbf{Probe B. Rhythmic groove selectivity (Ballroom Dataset).}
Triplets are formed from dance-music recordings, where positive and anchor share a dance class (e.g., both Tango) and the negative belongs to a different class.
The eight ballroom classes are defined by rhythmic signature (meter, BPM range, syncopation pattern) under near-constant big-band instrumentation, an unusually clean rhythm-only label.
$H_\text{rhy}$ reaches 88.0\%, the highest single-head accuracy across all three probes, while $H_\text{mel}$ drops to 55.2\%, a 32.8-point selectivity gap, and $H_\text{tim}$ sits at 67.2\%.
Crucially, $H_\text{rhy}$ exceeds raw MERT (78.0\%) by a full 10 percentage points despite the encoder being identical: training has bent the projection to align with rhythmic signature and away from competing acoustic cues.
This is the strongest evidence in the paper that the supervision signal alone, on top of a frozen backbone, can extract a perceptually defined factor that the backbone does not separate by default.

\textbf{Probe C. Multi-factor profile (Covers80).}
Covers80 pairs an original recording with a cover version of the same song.
Pairs are expected to share \emph{multiple} perceptual factors: melody is preserved by definition, but rhythm and tempo are typically preserved to some extent, since most covers are arrangements rather than reinterpretations.
Accordingly, both $H_\text{mel}$ (63.4\%) and $H_\text{rhy}$ (69.9\%) score above the MERT cosine baseline (66.1\%), while $H_\text{tim}$ (61.3\%) sits below it, consistent with the fact that cover identity is primarily a combined melodic and rhythmic phenomenon and optionally a timbral one.
The fact that $H_\text{rhy} > H_\text{mel}$ on Covers80 is, on first reading, surprising; on closer reading, it is a known property of the corpus. Most pairs in Covers80 are tribute or live arrangements, so groove is preserved more reliably than melody is recovered by a representation trained to be \emph{invariant} to instrumentation.
The score combination concat strategy (69.5\%) exceeds any single head, demonstrating that the three factor projections are complementary rather than redundant.

\subsection{Per-Class Selectivity Patterns}
\label{sec:perclass}
Diving deeper into the probe results from above, we see that on MUSDB18-HQ, the accuracy of $H_\text{tim}$ peaks on drums (90.7\%) and is weakest on vocals (70.1\%) where pitched articulation introduces timbral variability across singers; $H_\text{rhy}$ rises to 83.2\% on vocals, a known cross-factor artifact of the consistent rhythmic phrasing of pop vocal lines.
On the Ballroom Dataset, $H_\text{rhy}$ reaches 97.0\% on Tango, 94.4\% on Viennese Waltz, and 94.0\% on standard Waltz (three classes with maximally distinctive meters), and drops to 65.4\% on Rumba-International, where the three Rumba subclasses share an overlapping syncopation profile; the rhythm head degrades only on the subclass cut where the rhythmic signature is genuinely ambiguous.

\subsection{Score Fusion Strategies}
\begin{figure}[t]
  \centering
  \includegraphics[width=\columnwidth]{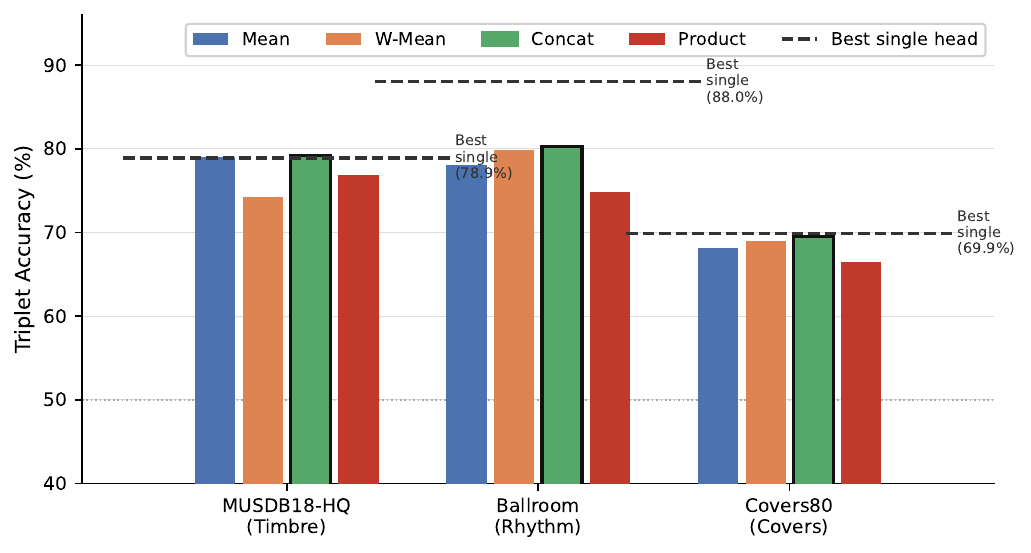}
  \caption{\textbf{Score fusion strategies vs.~best single head.}
    Triplet accuracy (\%) for four fusion strategies applied to all
    three MERIT heads simultaneously: unweighted mean (Mean),
    weighted mean (W-Mean, weights tuned on Covers80),
    $\ell_2$-normalised concatenation (Concat),
    and element-wise product (Product).
    The dashed line marks the \emph{best single head}---the highest accuracy
    achievable by any one factor head alone on that probe
    ($H_\text{tim}$ on MUSDB18-HQ; $H_\text{rhy}$ on Ballroom and Covers80).
    % Concat is the only strategy that consistently matches or surpasses the best individual head across all three probes.Chance level is 50\% (dotted line).
    }
  \label{fig:combo}
\end{figure}

Fig.~\ref{fig:combo} compares four fusion strategies against the per-probe best single head across all three probes. The \emph{best single head} baseline is the highest accuracy achieved by any one factor head in isolation for that probe: $H_\text{tim}$ (78.9\%) for MUSDB18-HQ, and $H_\text{rhy}$ (88.0\% and 69.9\%) for Ballroom and Covers80, respectively.
For Probe A and Probe C, the $\ell_2$-normalised 384-dimensional concatenation $[H_\text{mel};H_\text{rhy};H_\text{tim}]$ matches or exceeds any single head, confirming that the three projections are \emph{complementary} rather than redundant; element-wise product fares worst because a single near-zero factor score nullifies the others.

% \subsection{Score Combination Analysis}
% \label{sec:combo}

% A practical question is how to combine the three head outputs into a single retrieval score when the user has no prior over which factor matters.
% We compare four fusion strategies across all probes (Table~\ref{tab:probes}, lower block):
% (i)~unweighted mean of the three 128-dim vectors followed by $\ell_2$-normalisation;
% (ii)~weighted mean with cover-tuned weights $(w_\text{mel},w_\text{rhy},w_\text{tim})=(0.40,0.40,0.20)$;
% (iii)~$\ell_2$-normalised concatenation into a single 384-dim representation;
% (iv)~element-wise product of the three vectors.
% Concatenation is the most consistent winner: it tops the aggregate on Ballroom (80.3\%) and Covers80 (69.5\%), and on MUSDB18-HQ (no mixture) sits within noise of the max-pool variant (79.2\% concat vs 79.5\% max).
% Product-of-experts under-performs because any near-zero factor score nullifies the others, and the weighted mean is sensitive to the prior weights.
% The fact that concatenation, which preserves all three factor signals separately rather than collapsing them, beats the alternatives confirms that the three projections are \emph{complementary} rather than redundant; there is meaningful per-factor variance that fusion should preserve, not average away.

\begin{figure}[h!]
    \centering
    \includegraphics[width=1\linewidth,trim={0 20px 0 20px}, % Left Bottom Right Top
  clip]{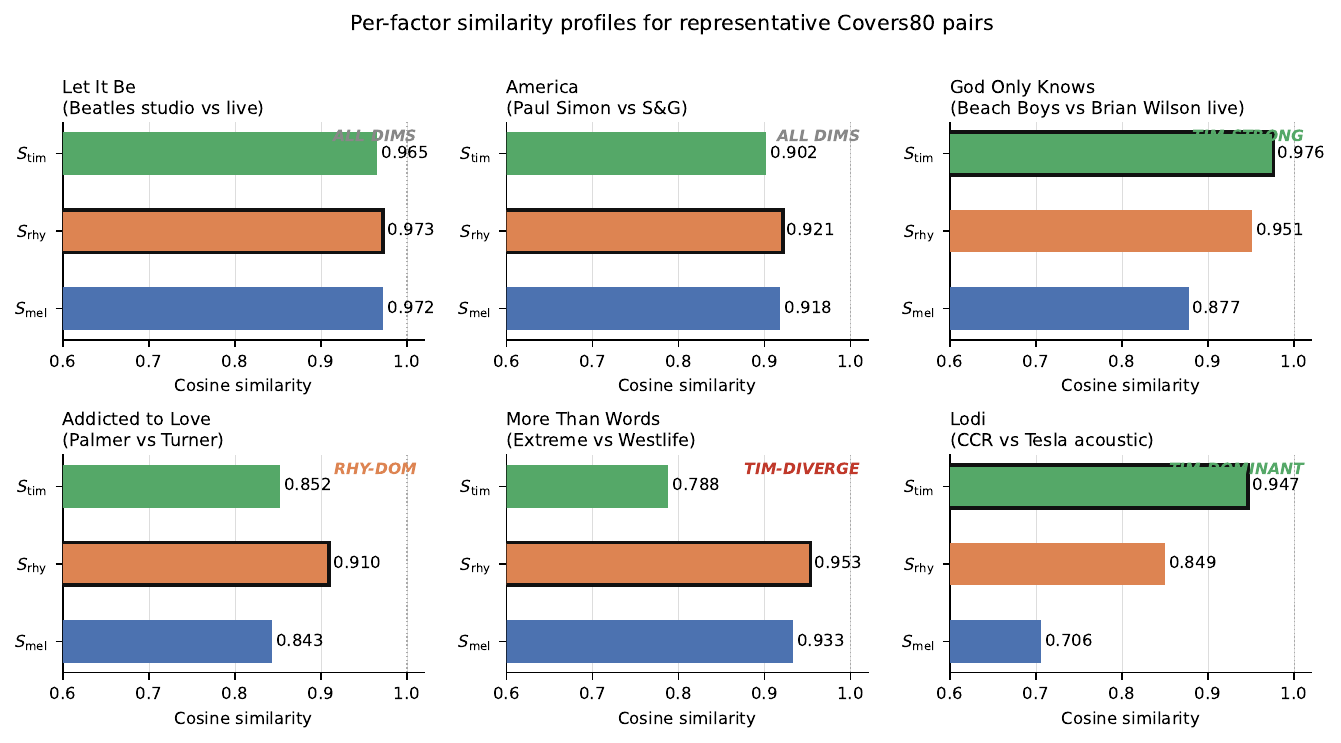}
  
    \caption{\textbf{Per-factor similarity profiles for Covers80 pairs.} Each group of bars shows the MERIT scores for one cover pair.}
    % \$(S_\text{mel},\,S_\text{rhy},\,S_\text{tim})$
    \label{fig:profiles}
\end{figure}

\subsection{Qualitative Factor Profiles}

Beyond aggregate accuracies, we look into individual pairs as case studies.
We extract per-factor similarities $(S_\text{mel}, S_\text{rhy}, S_\text{tim})$ from each trained head for three representative cover pairs in Covers80; these are reported in Fig.~\ref{fig:profiles}.
``Let It Be'' (Beatles studio vs Beatles live) scores near-identically on all three factors ($0.97$ each), as expected for a same-artist cover that preserves arrangement.
``More Than Words'' (Extreme vs Westlife) preserves melody and rhythm ($S_\text{mel}=0.93$, $S_\text{rhy}=0.95$) but diverges in timbre ($S_\text{tim}=0.79$), reflecting Westlife's pop-vocal arrangement against Extreme's acoustic-rock original.
``Lodi'' (Creedence Clearwater Revival vs Tesla's acoustic cover) inverts the pattern: timbre is the strongest match ($S_\text{tim}=0.95$) while melody is the weakest ($S_\text{mel}=0.71$), capturing the audible fact that Tesla's acoustic-jam arrangement reproduces CCR's instrumentation more faithfully than its melodic phrasing.
A monolithic similarity score cannot articulate this distinction.

\subsection{Layer-Attribution Analysis}

\label{sec:layer-attr}

\begin{figure}[t]
    \centering
    \includegraphics[width=1\linewidth, trim={0 20px 0 20px}, % Left Bottom Right Top
  clip]{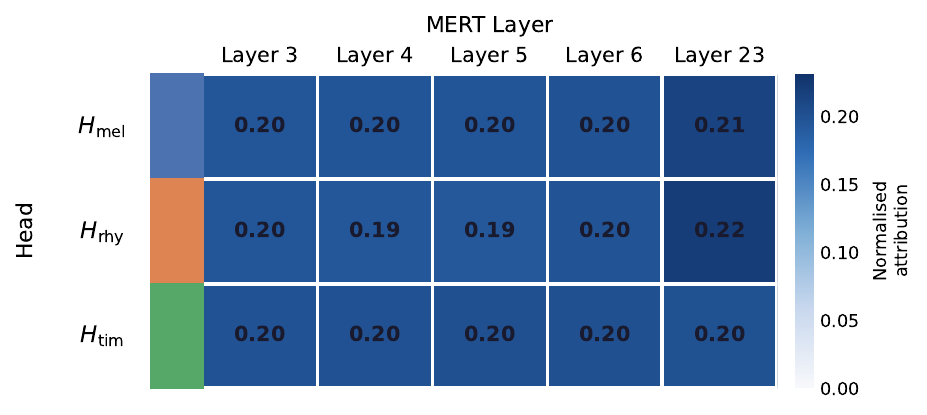}
    \caption{\textbf{MERT layer attribution per factor head.}
      Each cell shows the row-normalised Frobenius norm of the submatrix of $\mathbf{W}_1^f$ corresponding to that MERT layer,
      quantifying the fraction of the head's first-layer weight mass that attends to each depth.
      $H_\text{mel}$ concentrates on the deeper layer~23, consistent with pitch and melodic contour being encoded at higher abstraction levels.
      $H_\text{rhy}$ relies more on the shallower layers~3--6, which capture low-level temporal periodicity and rhythmic texture.
      $H_\text{tim}$ shows a broader distribution, reflecting the fact that timbral identity is expressed across multiple levels of spectral and acoustic abstraction.
      No explicit depth bias was imposed during training; the specialisation emerges solely from the factor-specific supervision signal.}
    \label{fig:layer_attribution}
\end{figure}

The specific backbone design admits a direct read-out of which MERT depths each head relies on: the first-layer weights $\mathbf{W}_1^f \in \mathbb{R}^{512\times 5120}$ partition along the input axis into five $512\times 1024$ submatrices, one per MERT layer, whose Frobenius norms quantify how strongly head $f$ attends to that layer.
Fig.~\ref{fig:layer_attribution} reports the row-normalised heatmap and shows that the supervision signal alone selects the depths in MERT that are diagnostic of each factor, consistent with prior probing studies that locate pitch and rhythm information at different depths in self-supervised audio encoders.

% \subsection{Margin Geometry and Effect Sizes}
% \label{sec:margin}
% Triplet accuracy is a binary summary of a continuous geometric quantity: the cosine-distance margin $m = d_{AN} - d_{AP}$ between negative and positive distances to the anchor.
% Reporting margins alongside accuracy clarifies the magnitude of the factor signal and not merely its sign.
% On the diagonals, $H_\text{mel}$ separates anchor--positive from anchor--negative pairs by a mean margin of $+0.788$ ($d_{AP}=0.155$, $d_{AN}=0.943$ on the unit cosine sphere) and $H_\text{rhy}$ by $+0.862$, both an order of magnitude larger than the raw-MERT margins on the same test sets ($+0.034$ on melody, $+0.048$ on rhythm).
% On the critical $H_\text{rhy}$-on-melody cell, the margin is not merely small but \emph{negative} ($-0.038$, with $d_{AP}=0.667 > d_{AN}=0.628$): the rhythm projection places melodically matched clips slightly farther apart than mismatched ones.
% Reporting these signed margins also disambiguates ``borderline'' cells: a 50\% accuracy with a positive margin centered near zero is qualitatively different from a 50\% accuracy with a negative margin, and only the latter constitutes evidence of suppression.

\section{Discussion}
The diagonal scores near 100\% in Table~\ref{tab:disentanglement} reflect supervision aligned with what a shallow MLP on multi-layer MERT can extract; the held-out test split is folder-disjoint from training, so this is not overfitting in the conventional sense.
A residual concern is that the within-pipeline test set could inherit JASCO-borne correlations that a head might exploit without learning the intended factor; the zero-shot probes directly address this on the evaluation side, since MUSDB18-HQ, the Ballroom Dataset, and Covers80 contain no JASCO audio and no MoisesDB stems, yet $H_\text{tim}$ and $H_\text{rhy}$ remain the dominant heads on their respective probes.
The Covers80 result, where $H_\text{rhy}$ slightly exceeds $H_\text{mel}$, reflects a known property of that corpus: tribute and live arrangements preserve tempo and groove nearly as faithfully as melody, and our $H_\text{mel}$ is trained to be invariant to instrumentation and rhythm, suppressing exactly the cues that single-vector cover detectors rely on. This work's claim is not that $H_\text{mel}$ is a state-of-the-art cover detector but that the per-factor profile reveals which dimension of similarity each pair preserves.

Certain limitations open up scope future work: the decomposition is restricted to melody, rhythm, and timbre, with harmony and dynamics requiring additional conditioning channels; timbre is operationalised at instrument-class granularity through MoisesDB labels, which under-resolves within-class variation (two acoustic guitars with different recording conditions are treated as identical positives); and JASCO's conditioning fidelity sets a ceiling on melody and rhythm supervision quality.

\section{Conclusion}

We presented \textbf{MERIT}, a representational framework that exposes melodic, rhythmic, and timbral similarity as three separable scores.
On three zero-shot probes, the intended head is the strongest factor head on instrument-class identity (MUSDB18-HQ) and on dance-style rhythmic signatures (Ballroom), and the cross-factor profile recovered on cover pairs (Covers80) matches the audible character of individual pairs.
As future work, we see two natural extensions: adding harmony as a fourth factor through chord-conditioned generation, and replacing the per-factor projection heads with a single multi-head transformer that allows the heads to share intermediate computation while remaining selectively supervised. The source code and pre-trained projection heads are publicly available at \url{https://github.com/AMAAI-Lab/MERIT}; the training triplets are available at \url{https://huggingface.co/datasets/amaai-lab/merit}.

\section{Acknowledgments}
This work has received funding from grant no. SUTD SKI 2021\_04\_06 and from MOE grant no. MOE-T2EP20124-0014

\section{AI Usage Statement}
We acknowledge the use of Gemini and ChatGPT for paraphrasing and grammar improvements.

% For BibTeX users:
\bibliography{ISMIRtemplate,amaai}

\end{document}